\documentclass[prd,a4paper,aps,epsfig,showpacs,twocolumn,floats]{revtex4}

\newcommand{\be}{\begin{equation}}
\newcommand{\ee}{\end{equation}}
\newcommand{\bea}{\begin{eqnarray}}
\newcommand{\eea}{\end{eqnarray}}

\begin{document}
\title{Speedy sound and cosmic structure}
\author{Jo\~ao  Magueijo}
\affiliation{
Theoretical Physics Group, Imperial College, London, SW7 2BZ}
\date{\today}
\begin{abstract}
{If the speed of sound were vastly larger in the early Universe
a near scale-invariant spectrum of density fluctuations could
have been produced even if the Universe did not submit to conventional 
solutions to the horizon problem. We examine how the mechanism works, 
presenting full mathematical solutions and their heuristics. 
We then discuss several concrete models based on scalar fields and 
hydrodynamical matter which realize this mechanism, but stress
that the proposed mechanism is more fundamental and general.}
\end{abstract}
\pacs{0000000}
\maketitle

\bigskip

{\it 1.Introduction} The fact that the large scales we observe today were ``out of touch'' 
in the early Universe is one of the greatest annoyances of standard 
cosmology. This ``horizon'' problem prevents a causal explanation for 
the observed features of our Universe,
which have to be added on ``by hand'' as initial
conditions. Nowhere is this more unsatisfactory than in relation to
the primordial density fluctuations that seed the structures
we observe today. We have measured these structures with tremendous 
accuracy; yet the primordial  fluctuations cannot be
explained in the standard scenario and remain  
God-given initial conditions. This weakness has motivated
many revisions to the standard Big Bang picture, 
notably inflation~\cite{infl}.

Any explanation for the initial spectrum of fluctuations
has to begin with a mechanism for causally connecting
vast scales at the cradle. 
But this is barely the beginning:  one then has to suggest a physical mechanism  
that would render these scales homogeneous but ``not quite''. In the detail
lies the hurdle: the inhomogeneities must be near scale-invariant,
i.e. look approximately the same on all scales, and have a well-defined amplitude, of the 
order of a part in 100,000. Herein lies  the
challenge for any structure formation scenario.

Solutions to the horizon problem rely on either accelerated expansion~\cite{infl},
a contracting phase followed by a bounce~\cite{ekp}, or a loitering early phase~\cite{hag}. 
An alternative was supplied by varying speed of
light (VSL) theories~\cite{vsl,jm,am,mach}: the idea that the early Universe operated under a much
larger maximal  speed, causally connecting everything. There are several 
implementations of 
this idea and structure formation has been considered in a few~\cite{mofstruc}. 
But in spite of these valiant efforts it is fair to say that 
VSL scenarios have so far failed to explain the cosmic structures.

In this paper we use a varying speed of sound as a proxy for VSL.
In theories containing two metrics~\cite{bim} the speed of massless matter
particles and the speed of gravity are different. This defines a frame where
gravity is unaffected, but ``light'' travels much faster.
We propose a new mechanism for producing scale-invariant
fluctuations, based on a varying speed of sound. The idea mimics that
of a varying speed of light. If the sound horizon shrinks, modes
that start off oscillating eventually freeze-out. The universe
is initially homogeneous apart from small quantum or thermal fluctuations.
These are left imprinted ``outside the horizon'' after the speed of sound
$c_s$ has decreased suitably.

A varying speed of sound has been considered in other guises before. 
We nod at $\kappa$-essence models, based on (scalar) field theories
with non-standard kinetic terms~\cite{kappa}. These have been used as inflationary
and quintessence models, but can be adapted to implement the varying 
speed of sound mechanism we advocate. The cuscuton model 
of~\cite{cusc1,cusc2} provides a related framework. 
We can also 
bypass scalar fields,
and simply regard the speed of sound $c_s$ and 
the equation of state $w$ as free phenomenological parameters~\cite{hu,jochen}. Even though VSL is our 
leading motivation here, we stress that many other approaches may be linked with the
conclusions in this paper. A VSL implementation, however, may be required
if additionally one wants to solve the flatness, entropy and the other
problems of Big Bang cosmology.

{\it 2.The mechanism} 
We shall first illustrate the mechanism in its simplest realization.
Suppose that gravity remains unmodified 
and that we 
restrict ourselves to expanding Universes with $w>-1/3$.
This is 
to avoid confusion with inflation and ekpyrotic scenarios
(although constructive alliances should be investigated). 
The new ingredient, now, is the assumption that the speed of sound 
is density dependent and diverges with conformal time like
$c_s\propto \eta^{-\alpha}$ (with $\alpha>0$; note that $\eta$ is positive
and increases from zero). A reparameterization in terms of the density 
will be examined later (with a striking result) but this is the most
suitable expression for a mathematical solution. Concrete models 
realizing this set up will be presented but we do not want to wed what follows 
to any one of them. 

Whether we employ a fluid or a scalar field the density fluctuations 
are described by a modified harmonic oscillator equation. This can
be written in terms of variables related to the Newtonian potential $\Phi$ or the
curvature perturbation $\zeta$, the so-called ``$u$'' and ``$v$''.
The equation for $v$ is~\cite{mukh,lidsey,garriga}:
\be\label{veq}
v''+\left[c_s^2 k^2 -\frac{z''}{z}\right]v=0
\ee
where $z \propto {a}/{c_s}$.  Thus $c_s$ appears in
two places: in the $k^2$ term responsible for sub-horizon oscillations
and in the variable mass term, in $z''/z$. The equivalent equation for $u$ has
a variable mass term that does not contain $c_s$. 

As with inflation
this equation can be exactly solved with Bessel functions, but we'll first examine
a WKB solution in order to establish the initial conditions. 
With $\alpha>1$  modes start inside the sound horizon but eventually
leave it (the horizon scale is set by $c_sk\eta\sim 1$, i.e.
$k\propto \eta ^{1-\alpha}$).
We can therefore initially ignore the term in $z''/z$. 
Even though the frequency $\omega=c_s k$ is changing, 
the WKB condition $\omega '  \ll \omega^2$ translates into 
$c_s k\eta\gg |\alpha|$, which is always satisfied early on.
This amounts to preserving the adiabatic invariants, i.e.
the number of quanta in a given mode is kept fixed
but not the total energy in that mode, 
which changes proportionally to the frequency.
The appropriately normalized WKB solution is therefore
\be\label{bc1}
v\sim\frac{e^{ik\int c_s d\eta}}{\sqrt{c_s k}}\sim 
\frac{e^{-\beta c_s k \eta}}{\sqrt{c_s k}}
\ee
where $\beta=1/(\alpha-1)>0$ and in the last $\sim$ we neglected a 
phase. We can also consider scenarios where $c_s$ is initially constant, 
then drops like a power law. The standard boundary condition may then be 
imposed in the constant $c_s$ phase and propagated using the WKB solution, 
with the same result.

As with the equivalent calculation in inflation, Eqn.~(\ref{veq}) 
can be transformed into a Bessel equation, with solutions:
\be
v=\sqrt{\beta\eta}(AJ_\nu(\beta c_sk\eta)+BJ_{-\nu}(\beta c_s k\eta))\; .
\ee
The order $\nu$ is given by 
$
\nu=\beta{\left(\alpha-\frac{3(1-w)}{2(1+3w)}\right)}
$
and $A$ and $B$ are $k$-independent numbers of order 1, so that the boundary
condition (\ref{bc1}) is satisfied. The spectrum left outside the horizon
is now easy to find. Since $c_s\eta$ is a decreasing function of time, 
the negative order solution is the growing mode, so that asymptotically we have:
\be\label{vout}
v\sim \frac{\sqrt{\beta \eta}}{(c_s k \eta)^{\nu}}\; .
\ee
Since the curvature fluctuation is related to $v$ by $\zeta=v/z$, its scale-invariance
($k^3\zeta^2=const$) requires $\nu=3/2$, i.e.
\be\label{alpha0}
\alpha=\alpha_0=6\frac{1+w}{1+3w}\; .
\ee
If we rephrase this requirement by writing $c_s$ in terms of the density
$\rho$ we conclude, interestingly, that $c_s\propto \rho$  for all $w$.
The spectrum can also be made red or blue depending
on whether $\alpha<\alpha_0$ or $\alpha>\alpha_0$, specifically
\be\label{nvac}
n_S-1=\beta (\alpha-\alpha_0)\; .
\ee 
We note that 
an infinitely fast transition ($\alpha\gg 1$) implies 
$n_S=2$ for all $w$. All of these considerations depend on the
sub-horizon normalization, here chosen to match a vacuum quantum state.

We can also work out the fluctuations' amplitude for near scale-invariant spectra.
Considering the curvature  $\zeta$, which ``freezes-in'' 
(i.e., is time independent outside the horizon, even with 
a variable $c_s$), it is found~\cite{garriga} 
after straightforward algebra: 
\be\label{ampcsrho}
k^3\zeta^2\sim \frac{(5+3w)^2}{1+w}\frac{\rho}{M^4_{Pl}c_s}\; .
\ee
Those acquainted with this expression (say, from inflation) will 
find here a good explanation for why $c_s\propto \rho$ leads 
to scale-invariance, even without inflation.
If we refine the $c_s$ law to $c_s=c_0(1+\rho/\rho_\star)$, (where $c_s\approx c_0$ 
at low-energy and $\rho_\star$ is the density that triggers
its divergence) we find:
\be\label{amp}
k^3\zeta^2\sim \frac{(5+3w)^2}{1+w}\frac{\rho_\star}{M^4_{Pl}}\sim 10^{-10}\; .
\ee
This forces the energy scale of the varying speed of sound phenomenon
to be a couple of orders of magnitude below the Planck scale, unless
$w\gg 1$ in which case it can be significantly lower.

This is not to say that this scenario is protected from 
the super-Planckian ``problem'' (cf.~\cite{hollands} for a similar 
affliction). Indeed, choosing $w=1/3$ to fix ideas, 
and taking into account the 
amplitude (\ref{amp}), straightforward algebra shows that the current Hubble
scale left the sound horizon when $H_0^{-1}=H^{-1}(\rho)c_s(\rho)a_0/a(\rho)$,
translating into $\rho/M_{Pl}\sim 10^{24}$, or  
$E \sim 10^6 M_{Pl}$. This can only be evaded with a very large $w\gg 1$,
leading to $E_\star <10^{-10}M_{Pl}$ so that the current Hubble scale 
does freeze-in for $E<M_{Pl}$; however for the normalization to be correct 
one would then need $w>10^{30}$. We would not discount 
more standard scenarios (with $w$ of order one), as this
``problem'' might be quite fictitious: 
it could well be that there is never a Planckian quantum gravity
phase.

Perhaps more interesting is the possibility that the initial conditions
are ``thermal'' rather than quantum vacuum fluctuations as assumed 
above~\cite{pedro}. The sub-horizon modes then have a spectrum 
obtained from the above by a multiplicative factor of $T_c/k$ 
(where $T_c=T a/c_s$ is a constant). This reflects the Rayleigh-Jeans  
limit of  the thermal occupation number $n(k)$ and implies a sub-horizon 
white-noise spectrum. This then propagates to the final result 
(Eqns.~\ref{vout}
or \ref{nvac}) leading to the prediction
\be
n_S-1=-\frac{\alpha_0-1}{\alpha-1}\; .
\ee
Scale-invariance is now ensured by a very fast phase 
transition ($\alpha\gg \alpha_0$, i.e. something close to a step-function).  
The spectrum is always red but can also be made arbitrarily close to flat.
The amplitude, if departures from scale-invariance are small,
can be found following some simple algebra, keeping track of all 
significant multiplicative constants, and is:
$
A^2=\alpha{\left(T_\star/M_{Pl}\right)}^3\sim 10^{-10}
$.
Thus these models do not suffer from a trans-Planckian problem.
With $\alpha\sim 100$ we have $\rho_\star/M_{Pl}^4\sim 10^{-16}$
and the current horizon scale leaves the horizon at a density
barely an order of magnitude higher than this.
If anything these scenarios may have the opposite problem: they
push $E_\star$ very low if the transition is extremely steep.
For example with $\alpha\sim 10^{50}$ the appropriate normalization
would require $T_\star \sim 1$Gev, with the current horizon scale leaving 
the horizon at around the same energy scale.
This might bring these scenarios within the reach of direct experimental test.
Perhaps the study of high energy thermal plasmas~\cite{nova} could
provide a direct measurement of $c_s$, placing the model under laboratory
test.

To conclude our presentation of the mechanism we note that
in general one might have both thermal and quantum contributions. 
In the absence of the usual inflationary justification for a vacuum
state the thermal argument may make more sense.
Also notice that in these scenarios the curvature $\zeta$ and the potential
$\Phi$ both freeze outside the horizon and have the same spectrum 
and normalization (within factors of order 1). This is to be contrasted with 
the pathologies known to plague some scenarios~\cite{gratton}.

{\it 3.Models} So far we have concentrated on investigating a new {\it mechanism} for producing 
scale-invariance and on its phenomenology. We specifically avoided 
wedding it to any ``model'' (although this has to be done to solve the flatness
and other Big Bang problems).
However we'll happily exhibit some possible models realizing this mechanism.
These should be merely seen as proof of concept.

For example, one can appeal to a scalar field $\phi$ endowed with non-trivial kinetic 
terms~\cite{kappa,kappa-ruth,vikman,bean}, 
a suitable potential to recreate constant $w$ (i.e. a scaling solution),
and a varying $c_s$. The Lagrangian for such models has the form
${\cal L}_0=K(X)-V(\phi)$ (with $X=\frac{1}{2}\partial_\mu\phi\partial^{\mu}\phi$).
From the stress energy tensor we find that 
$p=K-V$ and $\rho=2XK_{,X}-K+V$, 
i.e. $w$ depends on both $K$ and $V$. In contrast the speed of sound is given by:
\be\label{csK}
c_s^2=\frac{K_{,X}}{K_{,X}+2XK_{,XX}}
\ee
i.e. it depends only on $K$. For any power-law function $K(X)$ we can always find 
a power-law potential $V(\phi)$ that ensures a constant $w$ (scaling) solution. 
An example is: 
\be
{\cal L}_0=\sqrt {|X| X_\star}-m^2\phi^2\; ,
\ee
for which one can find a consistency relation between parameters $X_\star$ and $m$, as
well as initial conditions so that constant $w$ solutions, for any $w$, can be generated. 
As is well known this
model has an infinite speed of sound $c_s$ (cf. Eqn.~\ref{csK}): it's the so-called
cuscuton field~\cite{cusc1,cusc2}.  
But if we now add to ${\cal L}_0$ a new term ${\cal L}_1=X(X/X_\star)^{n-1}$
with $n< 1/2$ this term will be sub-dominant when $\rho\gg X_\star$, so that 
the initial, constant $w$ solution is still valid in this regime.
However $c_s$ will no longer be infinite, and using Eqn.~(\ref{csK}) we find
$
c_s^2\propto X^{\frac{1}{2}-n}\propto \rho^{1-2n}\; .
$
Thus for $n=-1/2$ we have $c_s\propto \rho$, as required for scale-invariance. 
Under thermal initial conditions the more extreme $\alpha\rightarrow
-\infty$ limit would have to be considered.
Note that the theory has to contain a cut-off at low 
$X$ ensuring that the kinetic term becomes linear in $X$ at low $X$ (or else 
the field should decay into normal matter).

This is only an illustration, but it highlights the limitations of standard 
thermodynamical arguments relating $c_s$ and $w$. For hydrodynamical
matter $w=p/\rho$ whereas $c_s^2=\delta p/\delta \rho$. 
If the $\delta$ in this expression is an adiabatic partial derivative and the background 
also evolves adiabatically~\cite{kodama,hu} we must have 
$w=-3\frac{\dot a}{a}(1+w)(c_s^2-w)$, which results from 
$w_{,\rho}=(c_s^2-w)/\rho$, itself a simple rearrangement of:
\be\label{constraintcsw}
c_s^2=\frac{d}{d\rho}(w\rho)\; .
\ee
These relations contradict the assumptions in our perturbation calculation
($w$ cannot be constant), however they are blatantly violated
by scalar fields endowed with a potential, as in the example just
presented, or
more prosaically in the cases of inflation or quintessence
(for which $c_s=1$ but $w$ can be anything). 

Once this is noted there is no reason not to consider general forms 
of matter with $c_s$ and $w$ regarded as independent variables~\cite{jochen}.
Even with thermalized matter the evolution may be
non-adiabatic~\cite{hu} either for the fluctuations, the background, or both,
thereby violating (\ref{constraintcsw}).
If the background evolves adiabatically there must be entropy production 
$p\Gamma=\delta p-c^2_{ad}\delta\rho$ 
in order to have a constant $w$ and a varying $c_s$. Specifically,
the conditions of our calculation are  met 
if $\Gamma=(\rho/\rho_\star)^2\delta$, which can be seen
as the appropriate ``non-adiabatic condition''.

It is also possible that a more complicated solution to (\ref{veq}), allowing for a 
varying $w$ satisfying (\ref{constraintcsw}) and using the  proposed  varying 
$c_s$ mechanism, still finds a niche for scale-invariance. 
This possibility is currently under consideration, for example 
in the context of adaptations of the Chaplygin gas~\cite{chapl}
(or generalizations of the work in~\cite{amend,piao}).
The mathematics of such models is considerably less straightforward,
but there is one case where a simple solution may be found. If the initial conditions are
thermal, as discussed above, then we should have a step function in $c_s$ in order
to obtain scale-invariance (or require $\alpha\gg 1$). Let this step 
in $c_s(\rho)$ happen at $\rho=\rho_\star$ and take $c_s=c_{-}$ (for
$\rho<\rho_\star$) into $c_s=c_{+}\gg c_{-}$ (for $\rho>\rho_\star$). 
Integration of Eqn.~(\ref{constraintcsw}) leads to
$
w=c_{+}^2+(c_{-}^2-c_{+}^2){\rho}/{\rho_\star}\approx c_+^2{\left(
1-{\rho}/{\rho_\star}\right)}
$
(for $\rho\ge \rho_\star$) showing that in this case we can ignore
variations in $w$ while $c_s$ is changing, enforcing the conditions for
our calculation while  complying 
with (\ref{constraintcsw}). Thus the calculations presented  do not need
to be modified if thermal initial conditions are used to obtain scale-invariance; 
in contrast with quantum initial conditions (and $c_s\propto \rho$), for which Eqn.~(\ref{constraintcsw}) 
implies $w\approx c_s^2/3$, so that a whole new calculation is warranted.

We note that causality complaints~\cite{caus,cusc1} are bound to be model 
dependent. They don't affect bimetric and cuscuton models (see
discussions in~\cite{caus,cusc1,kbim}), 
or more general $\kappa$-essence models if they're seen
as bimetric theories~\cite{kbim}. It {\it may} 
be that one generally needs to embed
$c_s>1$ in bimetric VSL in order to prevent causality
violations; but the instantaneous change required in thermal scenarios
could open up other possibilities. But it could also be that 
VSL is not required to solve the causality paradoxes~\cite{causreply}.

{\it 4.Conclusions} In summary we have revisited VSL scenarios with reference to
structure formation~\cite{vsl}  in the context of  what we
hope is a simpler framework: a varying speed of sound. 
In VSL's initial formulation~\cite{jm,am}
the idea was to have increased symmetry as the Universe cooled down, 
transitioning from a Galilean Universe (with infinite speed of light) 
to the near Lorentzian Universe we see today. Obtaining a well defined formulation
of such scenarios proved challenging, particularly under hard breaking of Lorentz invariance
(for instance the issue of gauge choice became a physical one, and arbitrariness
ruled). Machian scenarios, where the constants of nature evolve along with the Universe,
as a power of $a$ were also considered~\cite{mach,covvsl}.
It is ironic that, as shown in this paper, thermal and quantum initial conditions provide room for
a phase transition~\cite{jm,am} and a Machian scenario~\cite{mach}, respectively.

How can we understand our results heuristically? In inflation a suitable
heuristic can be obtained by noting that the curvature $\zeta=v/z$
freezes outside the horizon and that inside the horizon $v$ becomes a regular
Minkowski scalar field. The same is true in our scenario. In inflation 
one has to match free oscillating modes of the form $v\sim e^{i k\eta}/\sqrt {2 k}$
inside the horizon with $\zeta = F(k)\propto k^{-3/2}$ outside the horizon. 
The matching is done when $k\eta\sim 1$ and therefore requires $z\propto 1/|\eta|$
for scale-invariance: a near deSitter background. 
Here a similar argument can be made, but now inside the
horizon modes have the well-known WKB form $v\sim e^{ik \int c_s }/\sqrt {c_s k}$,
resulting from the variation in their frequency $\omega=c_s k$. These have
to be matched with  $\zeta = F(k)\propto k^{-3/2}$ outside the {\it sound} horizon,
and the matching done when $c_sk\eta\sim 1$. Thus we should have 
$a \sqrt{c_s}\propto 1/\eta$, resulting in the general condition (\ref{alpha0}). 
In both cases study of the two extreme regimes (large and small) followed by 
suitable matching is enough to infer the final result. (Those conversant with 
Eqn.~\ref{ampcsrho}---say, from inflation, or $\kappa$-essence---will also 
quickly see why $c_s\propto \rho$ leads to scale-invariance.)

Are there any striking observational differences between this mechanism and
inflation? Deviations from  $n_s= 1$ can be easily obtained in this
scenario, so  we do not expect striking differences in terms of scalar fluctuations.
However in the simple models we have considered the horizon problem remains unsolved for gravitational
waves  and so we do not expect any tensor modes. This is to be contrasted
with inflationary models with $c_s>1$ where the tensor to scalar ratio is actually 
enhanced~\cite{vikman}. This feature may be bypassed, say in more complex bimetric
theories, a matter we are currently  investigating. 
Concrete implementations, such as the bimetric VSL theory, have 
also to be considered if one wants to address 
the other problems of Big Bang cosmology.

I'd like to thank C. Armendariz-Picon, C. Contaldi, R. Brandenberger, 
P. Ferreira, A. Liddle, Y. Piao and A. Vikman for helpful comments.

\end{document}